# Fuel consumption elasticities, rebound effect and feebate effectiveness in the Indian and Chinese new car markets

Prateek Bansal
prateekb@nus.edu.sg
Department of Civil and Environmental Engineering, National University of Singapore, Singapore.

Rubal Dua
rubal.dua@kapsarc.org
King Abdullah Petroleum Studies and Research Center (KAPSARC), Saudi Arabia.

## Abstract

China and India, the world's two most populous developing economies, are also among the world's largest automotive markets and carbon emitters. To reduce carbon emissions from the passenger car sector, both countries have considered various policy levers affecting fuel prices, car prices and fuel economy. This study estimates the responsiveness of new car buyers in China and India to such policy levers and drivers including income. Furthermore, we estimate the potential for rebound effect and the effectiveness of a feebate policy. To accomplish this, we developed a joint discrete-continuous model of car choice and usage based on revealed preference survey data from approximately 8000 new car buyers from India and China who purchased cars in 2016-17. Conditional on buying a new car, the fuel consumption in both markets is found to be relatively unresponsive to fuel price and income, with magnitudes of elasticity estimates ranging from 0.12 to 0.15. For both markets, the mean segment-level direct elasticities of fuel consumption relative to car price and fuel economy range from 0.57 to 0.65. The rebound effect on fuel savings due to cost-free fuel economy improvement is found to be 17.1% for India and 18.8% for China. A revenue-neutral feebate policy, with average rebates and fees of up to around 15% of the retail price, resulted in fuel savings of around 0.7% for both markets. While the feebate policy's rebound effect is low – 7.3% for India and 1.6% for China– it does not appear to be an effective fuel conservation policy.

# 1. Introduction

While China is the world's largest emitter of greenhouse gases (GHGs) with the largest automobile market, India holds the third spot globally in GHG emission and fourth in vehicle sales (Gupta et al., 2018; Timperley, 2019; Dua et al. 2021). The transport sector in both countries contributes substantially to GHG emissions. As the third-largest emitter, India's transport sector accounted for 11% of the nation's $CO_2$ emissions in 2016 (Janssens-Maenhout et al., 2017). With a compound annual growth rate (CAGR) of 8.4%, India's overall transport $CO_2$ emission is forecasted to be one thousand megatonnes by 2030 (NITI Aayog and Rocky Mountain Institute, 2017a). Thus, understanding the determinants of transport fuel demand in the world's two most populous and rapidly developing economies has both local and global implications. Locally, vehicle emissions in China contribute around 12–36% of the total $NO_X$, 10.7% of the total $PM_{10}$, and 16.8% of the total $PM_{2.5}$ (Wang et al., 2019). According to the 2020 environmental performance index, China and India are ranked 120 and 168 out of 180 countries (EPI, 2020).

Within the transport sector, a further increase in the new car market size is expected for both countries, given their rapid rate of urbanization and rising income levels. Private vehicle ownership in India is increasing at a 10% compound annual growth rate (NITI Aayog and Rocky Mountain Institute, 2017b). Similarly, vehicle ownership per capita in Chinese cities is increasing at a rate of 10-25% per year (Darido et al., 2014).

To combat the current and projected increase in GHG emissions from the passenger car sector, both countries have adopted several fuel conservation policies. For instance, China first introduced fuel economy standards for light-duty passenger vehicles in 2005, followed by the implementation of a Corporate Average Fuel Consumption (CAFC) target in 2012 to achieve fuel consumption of 5 litres per 100 km by 2020. Moreover, China also introduced a subsidy scheme in 2010 to encourage the Chinese automobile firms to produce vehicles with engine displacement below 1.6 litres (Chen et al., 2020). Similarly, India adopted CAFC norms in 2017, which require automakers to reduce the fuel consumption of light-duty vehicles under kerb weight of 3,500 kilograms to below 130 grams per km $CO_2$ until 2022. Another policy measure being considered by policymakers around the world including NITI Aayog, the Indian Government's think tank, is a feebate policy, i.e. reducing the upfront cost of fuel-efficient vehicles and increasing the same for vehicles with lower fuel economy (NITI Aayog and Rocky Mountain Institute, 2017b; Dua et al., 2021; Sheldon and Dua 2020).

The effectiveness of the above-discussed fuel conservation policies mainly depends on the consequent change in car buyers' preferences to buy fuel-efficient cars and car usage. For instance, fuel economy standards are likely to increase the overall fleet fuel economy, but the extent to which the induced driving due to low operating cost offsets the fuel consumption savings depends on the combined fuel price elasticity of the car choice and usage. Such induced demand for travel due to a fuel conservation measure is referred to as the rebound effect (Frondel et al., 2012; Menon and Mahanty, 2015). Estimating car and fuel price elasticity of fuel consumption and corresponding rebound effects are thus critical for the ex-ante evaluation of a fuel conservation policy.

Several studies have explored the effectiveness of fuel conservation policies using combined car preference and usage data in developed countries such as the USA (Bento et al., 2009; Goldberg, 1998; Jacobsen, 2013; Small and Van Dender, 2007; West, 2004) and Japan (Fullerton et al., 2014), but applications of joint models are limited in developing country context. Tan et al. (2019) is the only study in the Chinese context that simultaneously analyses car usage and car type preferences to quantify the welfare effects of fuel taxes. However, they estimate separate models for car composition and vehicle kilometres travelled (VKT), leading to inconsistent parameter estimates (see Section 2.1 for a detailed discussion). The only study on joint modelling of Indian car buyers' preferences to travel and car types uses old data from the year 2010 (Chugh and Cropper, 2017). Other studies in the developing country context either rely on only modelling the preference for car types or vehicle fleet composition while ignoring the rebound effects due to induced car usage (Chen & Lin Lawell, 2020; Xiao and Ju, 2014; Yang and Tang, 2019) or use aggregate time-series data to derive elasticities (Dahl, 2012; Lin and Zeng, 2013).

We fill this gap by analysing national-level data on Chinese and Indian car buyers' socio-demographics, revealed preferences for car types, and car usage from 2016-17. This detailed data enable us to estimate a *theoretically-consistent* joint discrete-continuous mixed logit model of car buyers' preferences for car types and usage. Using the parameter estimates of the model, we derive new estimates of i) the fuel price and income elasticity of fuel consumption, ii) the effect of a purchase price reduction and fuel economy improvements on the fuel consumption while quantifying the rebound effects, iii) the effect of a revenue-neutral feebate policy on fleet fuel economy and fuel consumption. This study presents the first theoretically-consistent fuel consumption elasticities for the Chinese automobile market, and provides new insights related to the effectiveness of feebate policies in India and China. Finally, we present the first comprehensive comparison of the results from both emerging automobile markets.

The remainder of the paper is organised as follows. Section 2 reviews literature on estimating the effectiveness of fuel conservation policies and corresponding fuel consumption elasticity in China and India. Section 3 details the specification and estimation of the joint discrete-continuous mixed logit model. Section 4 describes data collection, summary statistics, and empirical model specifications. Parameter and elasticity estimates, along with the results of the revenue-neutral feebate policy simulation, are discussed in Section 5. Conclusions, implications and limitations are summarised in the final section.

## 2. Literature review

### 2.1 Chinese automobile market

The literature mainly consists of two sets of studies. The first set of studies relies on analysing aggregated time-series or panel data using reduced-form regression models while correcting for various endogeneity biases. A study by Lin and Zeng (2013) estimates the price and income elasticity of gasoline demand using annual gasoline consumption data of 30 provinces in China from 1997 to 2008. They use regional diesel prices and international crude oil prices as instrumental variables for gasoline prices. The ranges of intermediate-run price and income elasticities of fuel consumption are [-0.49, -0.196] and [1.01, 1.05], respectively. Using a monthly vehicle sales and characteristics data in a fixed-effect regression framework, Sun et al. (2016) show that the gasoline-pricing reform in 2008 led to an increase in new vehicle fleet fuel economy of 6.25%. In another study, He et al. (2017) investigate China's fuel demand system by analysing the panel data on vehicle population and utilisation of 306 cities in China from 2002 to 2012 using the Almost Ideal Demand System (AIDS) model. The estimated ranges of own-price and expenditure elasticity of fuel consumption for different vehicle models are [−1.22, −0.46] and [0.27, 1.24].

The second set of studies uses market- or consumer-level vehicle choice models. Xiao and Ju (2014) apply the popular market-level model of Berry, Levinson, and Pakes (BLP) to analyse the effect of consumption tax and fuel tax on vehicle sales and fuel consumption. Since they estimate the impacts of tax adjustments on both the demand and supply sides, they could estimate social welfare instead of consumer surplus. The results show that fuel taxes lead to reduced fuel consumption at the expense of social welfare, but the consumption tax has no significant effect on any of measures. Chen and Lin Lawell (2020) also use the BLP model to analyse the effect of fuel conservation policies on the vehicle market share and welfare in the Chinese automobile market. They show that China's CAFC standard that imposes restrictions on the sales-weighted average fuel consumption of the vehicles produced by a firm is inefficient. Instead, they strongly support the need for stringent fuel economy standards at the level of vehicle model. A few studies have also explored the effect of subsidies on fuel-efficient and new-energy vehicles. Yang and Tan (2019) use car registration data for five Chinese cities from 2011 to 2012 and estimate the BLP model to investigate the effectiveness of subsidies. They find that subsidies for fuel-efficient vehicles increase oil consumption and $CO_2$ emissions but still improve overall social welfare. However, subsidies for the new energy vehicles decrease gasoline consumption but raise $CO_2$ emissions and cause losses in social welfare. In another study, Sheldon and Dua (2020) explore cost-effectiveness and impact of the Chinese plug-in electric vehicle (PEV) subsidy program by estimating a consumer-level vehicle choice model using data on new vehicle purchases in the year 2017. They find that the cost of current PEV subsidy in China is $1.90 for an additional litre of gasoline saved. While the above-discussed studies ignore vehicle utilisation, Tan et al. (2019) simultaneously analyse vehicle choices and usage in the Chinese automobile market. They estimate fleet composition by applying the BLP model on market-level fleet composition and estimate vehicle usage using a fixed-effect regression model on panel data of individuals' VKT from 2009-10. They find that fuel price elasticity of vehicle usage is -0.59.

However, Tan et al. (2019) adopt a two-stage approach, i.e. they estimate vehicle demand and VKT models sequentially. This approach might lead to inconsistent parameters of both models in terms of different magnitude and sign (Chugh and Cropper, 2017). Leveraging household-level data on vehicle choice and usages, we address this limitation by jointly analysing both choices using a discrete-continuous mixed logit model. Along the lines of recent work by Bento et al. (2009) and Feng et al. (2013), we link vehicle demand and usage demand through Roy's identity, which leads to theoretically-consistent single set of parameter estimates. Readers are referred to Section 3 for details of model specification.

**2.2 Indian automobile market**

Only a few studies have explored the effectiveness of fuel conservation policies in India. Chugh et al. (2011) investigate the fuel economy valuation of Indian car buyers because the fuel economy standard is often preferred as a fuel conservation policy over fuel taxes, assuming that the consumers undervalue fuel economy. Using consumer-level data and a hedonic price approach, they find no strong evidence to support the assumption that Indian car buyers undervalue fuel economy. Bansal et al. (2021) also find similar results by analysing the preferences of Indian consumers for two-wheelers. They find that Indian two-wheeler buyers are not myopic, i.e. most of them use a discount rate of 10% or below to find the present value of future operating cost at the time of two-wheeler purchase. However, none of these studies accounts for the vehicle usage and rebound effects. Menon and Mahanty (2015) estimate such rebound effects by developing a system-dynamic simulation to analyse the effectiveness of alternative energy policies in conjunction with energy efficiency improvements in India. They find the rebound effects of these policies in terms of more car trips, but the simulation-based analysis is sensitive to assumptions on model parameters.

Chugh and Cropper (2017) present the most comprehensive analysis of the Indian automobile market by analysing household-level car choice and usage data. They deliver the first estimate of price and income elasticity of fuel consumption for the Indian market by applying the theoretically-consistent discrete-continuous model that we use in this study. However, their analysis uses old data from the year 2010 and substantially hinges on the differences in the retail price of diesel and petrol. Focusing on the duality of the automobile market was meaningful in 2010 because diesel has been cheaper than petrol historically due to higher subsidies. However, government-regulated petrol and diesel prices in India were directly linked to the international market rates in 2010 and 2014, leading to only small differences in diesel and petrol prices. For instance, the retail price of diesel climbed to $1.118 per litre, surpassing petrol price of $1.117 per litre petrol in the national capital in June 2020 (Bhardwaj, 2020). Such radical changes in fuel price norms are unlikely to encourage buyers to pivot their preferences on the minor temporary differences in prices of diesel and petrol, barring differences in vehicle fuel economy. This hypothesis is supported by a sharp drop in the market share of diesel vehicles from 58% in 2013 to 17% in 2021 (Rampal, 2021). Furthermore, the higher upfront price premium on diesel cars has been further exacerbated by the implementation of Bharat Stage-VI emission norms. The prohibitively high cost of upgrading diesel engines to meet the new emission norms has caused leading Indian carmakers including Maruti Suzuki to stop making diesel cars (Sasi, 2019). With these considerations, we use the household-level vehicle choice and usage data from the year 2017, and do not distinguish between vehicle alternatives based on fuel types. Along with new estimates of price and income elasticity of fuel consumption together with potential for rebound, we also provide new insights on the feasibility of feebate policies in the Indian context.

## 3. A model of new car purchases and usage

While joint modelling of household's preferences for energy products and their usage has been central to energy demand estimation since seminal work by Dubin and McFadden (1984), two-step approach has been popular until recent work by Bento et al. (2009) and Feng et al. (2013). These recent studies use Roy's identity in a static utility maximization framework to derive VKT and achieve single set of parameters for both choice models. This specification allows the simultaneous estimation of both choice models using full information maximum likelihood estimation. Both sub-models and the estimators are discussed below in detail.

### 3.1 Vehicle choice

According to Bento et al. (2009), the utility of household $i$ conditional on choosing car $j$ from $J$ available cars is

$$v_{ij} = u_{ij} + \varepsilon_{ij} = -\frac{1}{\beta_i}e^{-\beta_i(y_i - r_j) - \gamma X_{ij} - \eta_i} - \frac{1}{\alpha_i}e^{\alpha_i p_j} + \varepsilon_{ij}, \quad (1)$$

where $y_i$ is the annual income of household $i$, $r_j$ is the annualized purchase price or rental cost of car $j$, $X_{ij}$ includes car-specific, household-specific, and interaction of both characteristics, $p_j$ is the operation cost of car $j$ for one kilometre, and $\eta_i$ is an unobserved normally-distributed taste for driving with mean zero and standard deviation $\sigma$. We consider $\alpha_i$ and $\beta_i$ to follow a parametric distribution with parameters $\boldsymbol{\omega_\alpha}$ and $\boldsymbol{\omega_\beta}$, respectively. To maintain computational tractability of the model, we assume homogeneous sensitivity of utility relative to other household and car characteristics (i.e., constant $\boldsymbol{\gamma}$ across households). The common set of identified parameters in the systematic utility are $\boldsymbol{\Theta} = \{\boldsymbol{\gamma}, \boldsymbol{\omega_\alpha}, \boldsymbol{\omega_\beta}, \sigma\}$, and it follows a joint probability density function $f(\boldsymbol{u}|\boldsymbol{\Theta})$. Considering that $\varepsilon_{ij}$ is an independent and identically distributed idiosyncratic error term that follows Type 1 Extreme Value distribution with location parameter zero and scale $\mu$, the probability of choosing car $j$ by household $i$ is

$$P_{ij} = \int \frac{e^{u_{ij}/\mu}}{\sum_{k=1}^{J} e^{u_{ik}/\mu}} f(\boldsymbol{u}|\boldsymbol{\Theta}) d\boldsymbol{u}. \quad (2)$$

### 3.2 Driving distance

The annual VKT can be obtained from Equation 1 using Roy's identity:

$$KM_{ij} = -\frac{\partial v_{ij}/\partial p_j}{\partial v_{ij}/\partial y_i} = e^{\beta_i(y_i - r_j) + \gamma X_{ij} + \alpha_i p_j + \eta_i}, \quad (3)$$

We note that equations 1 and 3 have the same set of common parameters. Considering unobserved heterogeneity in the systematic utility, the expected value of annual driving distance is:

$$\mathbb{E}(\ln KM_{ij}) = \int [\beta_i(y_i - r_j) + \gamma X_{ij} + \alpha_i p_j + \eta_i] f(\boldsymbol{u}|\boldsymbol{\Theta}) d\boldsymbol{u}. \quad (4)$$

Considering that $\ln KM_{ij}$ takes standard linear regression form with normally-distributed error $\eta_i$ (i.e., taste of driving), the likelihood of observing $\widetilde{KM}_{ij}$ conditional on the household $i$ buys car $j$ is

$$\ell(\widetilde{KM}_{ij}|\mathbb{I}_{ij}=1) = \frac{1}{\sigma\sqrt{2\pi}} e^{-\frac{[\ln \widetilde{KM}_{ij} - \mathbb{E}(\ln \widetilde{KM}_{ij})]}{2\sigma^2}}, \quad (5)$$

where $\mathbb{I}_{ij}$ is an indicator which takes value 1 if household $i$ buys car $j$, else it is zero, and $\mathbb{E}(\ln \widetilde{KM}_{ij}) = \mathbb{E}[\beta_i(y_i - r_j) + \gamma X_{ij} + \alpha_i p_j]$, i.e. equation 4 without taste of driving as its expected value is zero.

### 3.3 Estimation strategy

The joint likelihood is the product of the likelihood of household $i$ to buy car $j$ and the likelihood of driving that car for $\widetilde{KM}_{ij}$ annually. Thus, the full information likelihood $\mathcal{L}$ and loglikelihood $\mathcal{LL}$ of a sample of $N$ households are

$$\mathcal{L}(\Theta, \mu) = \prod_{i=1}^{N} \prod_{j=1}^{J} [P_{ij} \ell(\widetilde{KM}_{ij}|\mathbb{I}_{ij}=1)]^{\mathbb{I}_{ij}}. \quad (6)$$

$$\mathcal{LL}(\Theta, \mu) = \sum_{i=1}^{N} \sum_{j=1}^{J} \mathbb{I}_{ij} \ln[P_{ij} \ell(\widetilde{KM}_{ij}|\mathbb{I}_{ij}=1)]. \quad (7)$$

Since the loglikelihood does not have a closed-form expression, we approximate the loglikelihood through simulation using a quasi-Monte Carlo method (Bansal et al., 2021). We also incorporate household-level weights in the analysis. The resulting weighted and simulated full information loglikelihood $\mathcal{WLL}$ of the sample is

$$\mathcal{WLL}(\Theta, \mu) = \sum_{i=1}^{N} \sum_{j=1}^{J} \mathbb{I}_{ij} \left[ w_i \ln \left( \frac{1}{R} \sum_{r=1}^{R} P_{ijr} \ell(\widetilde{KM}_{ijr}|\mathbb{I}_{ij}=1) \right) \right], \quad (8)$$

where $w_i$ is weight for household $i$, $P_{ijr}$ and $\widetilde{KM}_{ijr}$ are respective quantities for $r^{th}$ draw from $f(u|\Theta)$, and $R$ is the number of shifted and shuffled Halton draws. We estimate the joint discrete-continuous model by maximising the weighted simulated loglikelihood by writing our own MATLAB code with analytical gradient of the loglikelihood. We compute standard error of parameters using a robust sandwich estimator. To compute the standard errors of elasticity estimates, we take 100 draws from the asymptotic normal distribution of the estimator with mean as point estimates and robust hessian-gradient-based covariance matrix as the covariance. For each draw, elasticity value is calculated and standard deviation of elasticity estimates across 100 draws is reported as the standard error.

## 4. Data and model specification

### 4.1 Data for China and India

We use the initial quality survey (IQS) data collected in India and China by J.D. Power, a global leader in automotive marketing research. IQS data provide manufacturers with consumers' feedback and remain an industry benchmark for assessing new vehicle quality since 1987. New car buyers with ownership of two to six months, who purchased a car between November 2016 and July 2017 for personal use, were interviewed. The dataset consists of attributes of the purchased vehicle such as brand (make), model, segment (body type), fuel economy, and purchase price. This dataset also contains the delivery date, months of ownership, mileage on the vehicle, and various demographic characteristics of vehicles buyers (e.g., income, gender, and age). J.D. Power also provided us with aggregate sales data of cars to assess how well the individual-level sample represents the market of interest. We also have access to data collected by JATO for China on other vehicle characteristics (e.g., kerb weight and vehicle dimensions) at the model level.

Sales-weighted mean of segment-level car characteristics for China and India are presented in Tables 1 and 2, respectively. The data from China and India consist of 234 and 81 models. The summary statistics indicate that the sales-weighted fleet fuel economy of India is much higher than that of China (16.38 vs. 12.32 km/litre) because sport utility vehicles (SUVs) constitute a substantial share in the Chinese market, but compact and midsize segments dominate the Indian fleet with around 55% market share. As expected, Table 3 shows a similar and highly negative correlation of fuel economy with engine displacement and other vehicle size indicators (e.g., volume and weights) in both markets.

Tables 4 and 5 present segment-level demographic distribution for China and India. India and China samples have 7894 and 8951 households[1], respectively. Interestingly, 42.4% of new car buyers are female in China (with higher preferences for compact basic and luxury segments), compared to just 5% in India. The household size of Indian car buyers is 4.90, compared to 3.26 for Chinese car buyers. The average car buyer in China is 3.27 years younger than an Indian car buyer. The Annual VKT of both market buyers is close to 14,000 kilometres, with a much lower VKT of compact or mini segments as those might not be used for inter-city travel due to safety and comfort-related concerns. On average, the annual income of an Indian car buyer is much lower than that of a Chinese car buyer ($10,626 vs. $28,268, using 2017 conversion rates). As expected, luxury and premium car segments attract buyers with much higher income and higher ownership of cars in both markets.

The population of car buyers is not well defined because only a fraction of the population can be assumed to be willing and financially capable to buy new cars. Therefore, we adopt the choice-based (instead of exogenous demographics-based) sampling weights to get consistent parameter estimates. We compute sampling weights of each make-model such that choice proportions in the sample are the same as the actual sales proportions. The sampling weight ranges of India and China are [0.21, 8.18] and [0.14, 5.88], respectively.

---

[1] We take a random subsample of 40% observation from the originally-collected China data to make the estimation computationally tractable. With 234 alternatives and over 100 parameters, the estimation time of one model specification is around 50 hours for the considered China subsample. Since we use choice-based sampling weights, the randomly-selected subsample does not affect the consistency of the estimator.

Table 1: Sales-weighted mean of vehicle attributes for China market (standard deviations are in parenthesis).

| Segment | Fuel economy (km/litre) | Purchase price ($10^5$ CNY) | Engine displacement (litres) | Length x width x height ($10^{-10}$ mm$^3$) | Kerb weight ($10^{-3}$ kg) | Number of models | Market share |
|---|---|---|---|---|---|---|---|
| Compact | 14.74(0.78) | 0.53(0.08) | 1.36(0.07) | 1.03(0.04) | 1.05(0.05) | 3 | 1.63% |
| Compact Basic | 14.61(0.13) | 0.53(0.09) | 1.42(0.02) | 1.00(0.02) | 1.00(0.03) | 2 | 0.16% |
| Compact Luxury | 11.32(0.74) | 2.92(0.52) | 1.87(0.17) | 1.22(0.06) | 1.57(0.11) | 7 | 2.63% |
| Compact Luxury SUV | 11.03(0.49) | 3.13(0.47) | 1.77(0.12) | 1.31(0.06) | 1.64(0.09) | 6 | 1.42% |
| Compact Mini | 14.38(0.00) | 1.42(0.00) | 0.99(0.00) | 0.68(0.00) | 0.92(0.00) | 1 | 0.07% |
| Compact MPV | 13.31(0.25) | 0.63(0.10) | 1.49(0.05) | 1.36(0.08) | 1.31(0.09) | 8 | 5.17% |
| Compact SUV | 12.02(0.64) | 1.21(0.25) | 1.63(0.15) | 1.36(0.10) | 1.47(0.12) | 34 | 11.85% |
| Compact Upper | 14.33(0.55) | 0.85(0.02) | 1.46(0.02) | 1.05(0.04) | 1.09(0.02) | 6 | 2.88% |
| Large Luxury | 10.12(0.00) | 10.15(0.00) | 2.48(0.00) | 1.50(0.00) | 1.97(0.00) | 1 | 0.10% |
| Large Luxury SUV | 9.06(0.23) | 8.57(0.75) | 2.78(0.23) | 1.69(0.04) | 2.23(0.04) | 5 | 0.66% |
| Large MPV | 9.54(0.30) | 2.81(0.25) | 2.35(0.04) | 1.64(0.12) | 1.88(0.03) | 3 | 1.05% |
| Large SUV | 10.03(0.61) | 2.71(0.71) | 2.07(0.35) | 1.60(0.10) | 1.89(0.14) | 12 | 4.16% |
| Midsize | 13.15(0.44) | 1.21(0.18) | 1.55(0.06) | 1.21(0.03) | 1.29(0.04) | 25 | 20.46% |
| Midsize Basic | 13.74(0.55) | 0.85(0.11) | 1.51(0.04) | 1.17(0.05) | 1.20(0.09) | 24 | 10.60% |
| Midsize Luxury | 10.50(0.44) | 4.20(0.63) | 2.08(0.05) | 1.38(0.03) | 1.80(0.07) | 6 | 2.37% |
| Midsize Luxury SUV | 10.01(0.31) | 4.40(0.70) | 2.02(0.07) | 1.49(0.02) | 1.90(0.04) | 7 | 2.01% |
| Midsize MPV | 11.26(0.82) | 1.55(0.53) | 1.65(0.21) | 1.56(0.26) | 1.65(0.20) | 3 | 0.46% |
| Midsize SUV | 11.08(0.46) | 1.73(0.41) | 1.80(0.21) | 1.43(0.06) | 1.60(0.09) | 27 | 12.37% |
| Midsize Upper | 11.27(0.45) | 2.10(0.25) | 1.83(0.18) | 1.33(0.03) | 1.56(0.07) | 16 | 6.51% |
| Midsize Upper Economy | 12.57(0.37) | 1.54(0.21) | 1.54(0.10) | 1.24(0.05) | 1.39(0.07) | 12 | 5.04% |
| Mini Van | 14.15(0.39) | 0.37(0.04) | 1.24(0.10) | 1.14(0.06) | 1.00(0.05) | 3 | 0.77% |
| Small SUV | 12.95(0.73) | 0.81(0.12) | 1.55(0.09) | 1.29(0.12) | 1.33(0.11) | 23 | 7.63% |

Table 2: Sales-weighted mean of vehicle attributes for Indian market (standard deviations are in parenthesis).

| Segment | Fuel economy (km/litre) | Purchase price ($10^6$ INR) | Engine displacement (litres) | Number of models | Market share |
|---|---|---|---|---|---|
| Luxury | 13.08(1.30) | 4.89(0.65) | 2.16(0.18) | 7 | 0.37% |
| Upper Compact | 16.16(0.68) | 0.65(0.05) | 1.23(0.04) | 8 | 8.22% |
| MUV or MPV | 15.43(1.45) | 1.09(0.34) | 1.97(0.51) | 9 | 8.64% |
| Entry Midsize | 16.51(0.66) | 0.76(0.05) | 1.27(0.09) | 10 | 12.91% |
| Entry Compact | 17.58(0.74) | 0.39(0.04) | 0.83(0.10) | 6 | 11.53% |
| Compact | 17.04(0.64) | 0.50(0.04) | 1.03(0.07) | 8 | 14.90% |
| Premium Compact | 16.68(0.73) | 0.74(0.05) | 1.23(0.03) | 6 | 16.66% |
| SUV | 15.81(1.67) | 1.19(0.35) | 1.56(0.34) | 14 | 15.85% |
| Premium SUV | 13.10(1.52) | 3.05(0.01) | 2.73(0.22) | 2 | 0.95% |
| Midsize | 16.23(0.86) | 1.09(0.06) | 1.47(0.10) | 6 | 4.76% |
| Premium Midsize | 14.71(1.54) | 1.92(0.08) | 1.75(0.04) | 3 | 0.27% |
| Van | 15.21(0.19) | 0.39(0.06) | 1.02(0.20) | 2 | 4.94% |

**Table 3:** Correlation between car-specific attributes.

|  | Fuel economy | Purchase price | Engine displacement | Length x width x height | Kerb weight |
|---|---|---|---|---|---|
| **China** | | | | | |
| Fuel economy | 1 | | | | |
| Purchase price | -0.70 | 1 | | | |
| Engine displacement | -0.81 | 0.79 | 1 | | |
| Length x width x height | -0.80 | 0.51 | 0.68 | 1 | |
| Kerb weight | -0.93 | 0.77 | 0.82 | 0.86 | 1 |
| **India** | | | | | |
|  | Fuel economy | Purchase price | Engine displacement | | |
| Fuel economy | 1 | | | | |
| Purchase price | -0.68 | 1 | | | |
| Engine displacement | -0.78 | 0.58 | 1 | | |

**Table 4:** Mean socio-demographics of Chinese car buyers (standard deviations are in parenthesis).

| Segment | Age (years) | Female (%) | Number of cars | Family size | Annual income ($10^5$ CNY) | Annual kilometres | Market share |
|---|---|---|---|---|---|---|---|
| Compact | 32.81(7.20) | 0.44(0.50) | 1.09(0.28) | 3.03(0.79) | 1.35(0.08) | 13,690(7,412) | 1.63% |
| Compact Basic | 32.17(7.01) | 0.67(0.48) | 1.21(0.51) | 3.17(0.64) | 1.29(0.06) | 10,150(4,600) | 0.16% |
| Compact Luxury | 33.62(5.94) | 0.45(0.50) | 1.26(0.44) | 3.20(0.85) | 2.82(0.09) | 13,892(6,976) | 2.63% |
| Compact Luxury SUV | 33.03(5.71) | 0.52(0.50) | 1.30(0.55) | 3.13(0.80) | 2.92(0.08) | 13,192(6,170) | 1.42% |
| Compact Mini | 29.17(4.78) | 0.50(0.52) | 1.25(0.45) | 3.17(0.83) | 2.19(0.08) | 12,578(8,949) | 0.07% |
| Compact MPV | 33.37(6.64) | 0.22(0.42) | 1.06(0.24) | 3.52(0.94) | 1.48(0.09) | 14,286(7,025) | 5.17% |
| Compact SUV | 32.65(6.43) | 0.40(0.49) | 1.08(0.31) | 3.20(0.84) | 1.76(0.08) | 13,433(6,563) | 11.85% |
| Compact Upper | 31.37(5.70) | 0.49(0.50) | 1.05(0.26) | 3.16(0.76) | 1.60(0.08) | 13,284(6,112) | 2.88% |
| Large Luxury | 32.86(7.15) | 0.57(0.53) | 1.57(0.53) | 2.71(0.76) | 4.50(0.08) | 13,137(6,803) | 0.10% |
| Large Luxury SUV | 35.68(6.42) | 0.34(0.47) | 1.57(0.78) | 3.31(0.86) | 3.87(0.09) | 14,324(7,409) | 0.66% |
| Large MPV | 34.92(6.65) | 0.27(0.44) | 1.22(0.41) | 3.44(0.88) | 2.54(0.09) | 14,343(7,388) | 1.05% |
| Large SUV | 34.46(6.82) | 0.36(0.48) | 1.18(0.54) | 3.26(0.96) | 2.44(0.10) | 14,161(6,906) | 4.16% |
| Midsize | 32.49(6.38) | 0.45(0.50) | 1.06(0.24) | 3.17(0.84) | 1.66(0.08) | 13,181(6,540) | 20.46% |
| Midsize Basic | 32.28(7.10) | 0.42(0.49) | 1.07(0.26) | 3.20(0.88) | 1.44(0.09) | 13,010(6,433) | 10.60% |
| Midsize Luxury | 34.81(6.48) | 0.43(0.50) | 1.27(0.50) | 3.29(0.79) | 3.12(0.08) | 14,048(6,651) | 2.37% |
| Midsize Luxury SUV | 35.19(7.05) | 0.39(0.49) | 1.40(0.60) | 3.32(0.86) | 3.22(0.09) | 14,688(7,472) | 2.01% |
| Midsize MPV | 34.39(7.33) | 0.35(0.48) | 1.04(0.20) | 3.39(1.11) | 2.04(0.11) | 12,332(5,307) | 0.46% |
| Midsize SUV | 33.14(6.60) | 0.38(0.48) | 1.10(0.35) | 3.21(0.82) | 1.91(0.08) | 13,959(6,818) | 12.37% |
| Midsize Upper | 33.15(6.67) | 0.43(0.49) | 1.11(0.33) | 3.20(0.81) | 2.05(0.08) | 13,904(6,847) | 6.51% |
| Midsize Upper Economy | 32.92(6.82) | 0.42(0.49) | 1.07(0.32) | 3.14(0.80) | 1.85(0.08) | 13,934(7,196) | 5.04% |
| Mini Van | 31.75(6.55) | 0.19(0.39) | 1.06(0.24) | 3.17(0.73) | 1.43(0.07) | 14,486(9,115) | 0.77% |
| Small SUV | 32.63(7.02) | 0.40(0.49) | 1.07(0.31) | 3.20(0.84) | 1.53(0.08) | 13,395(6,804) | 7.63% |
| Sample | 33.08(6.92) | 0.42(0.49) | 1.13(.39) | 3.26(0.88) | 1.91(1.05) | 13,941(7271) | |

**Table 5:** Mean socio-demographics of Indian car buyers (standard deviations are in parenthesis).

| Segment | Age (years) | Female (%) | Number of cars | Family size | Annual income ($10^6$ INR) | Annual kilometres | Market share |
|---|---|---|---|---|---|---|---|
| Luxury | 39.19(8.78) | 0.05(0.23) | 2.38(0.76) | 4.97(1.19) | 1.35(0.24) | 14,021(10,141) | 0.37% |
| Upper Compact | 36.81(9.63) | 0.06(0.24) | 1.14(0.45) | 4.66(1.16) | 0.66(0.30) | 12,983(10,468) | 8.22% |
| MUV or MPV | 37.08(8.45) | 0.03(0.17) | 1.30(0.59) | 5.27(1.35) | 0.76(0.33) | 16,089(11,790) | 8.64% |
| Entry Midsize | 36.38(8.77) | 0.03(0.18) | 1.16(0.46) | 4.83(1.29) | 0.68(0.29) | 14,034(11,057) | 12.91% |
| Entry Compact | 36.09(9.91) | 0.07(0.25) | 1.16(0.44) | 4.73(1.36) | 0.56(0.28) | 11,939(9,935) | 11.53% |
| Compact | 36.63(9.90) | 0.06(0.25) | 1.14(0.47) | 4.79(1.37) | 0.62(0.29) | 11,754(8,860) | 14.90% |
| Premium Compact | 34.78(9.10) | 0.07(0.25) | 1.19(0.52) | 4.82(1.28) | 0.69(0.34) | 13,484(10,217) | 16.66% |
| SUV | 36.87(8.42) | 0.03(0.16) | 1.32(0.61) | 5.05(1.33) | 0.77(0.35) | 14,644(10,182) | 15.85% |
| Premium SUV | 37.80(8.84) | 0.07(0.25) | 1.87(0.86) | 5.48(1.43) | 1.19(0.35) | 16,088(9,953) | 0.95% |
| Midsize | 36.44(8.49) | 0.05(0.23) | 1.29(0.58) | 4.86(1.11) | 0.78(0.35) | 13,951(9,616) | 4.76% |
| Premium Midsize | 38.14(10.45) | 0.03(0.16) | 1.38(0.72) | 4.81(0.97) | 0.89(0.38) | 13,256(8,635) | 0.27% |
| Van | 36.92(9.28) | 0.01(0.08) | 1.16(0.57) | 5.16(1.22) | 0.53(0.28) | 12,942(9,499) | 4.94% |
| Sample | 36.35(0.09) | 0.05(0.21) | 1.23(0.55) | 4.90(1.32) | 0.69(0.33) | 14,037(10985) | |

### 4.2 Model specification

We convert the purchase price of a car to annualized rental price. We use an inflation-adjusted interest rate of 8.5% and an expected car life of 18 years for the Indian market (Chugh and Cropper, 2017). Similarly, for the Chinese market, we use an inflation-adjusted interest rate of 8% (S&P Global, 2019) and an average life span of 14.5 years (Hao et al., 2011b). The unit operating cost is obtained as the ratio of fuel price and fuel economy. We use the average 2017 fuel price of 6 CNY ($0.888) per litre for China (Ou et al., 2020) and INR 59 ($0.9086) per litre for India (ZeeBiz, 2017)[2]. Since fuel price remains the same for all models, the variation in operating cost arises from heterogeneity in fuel economy across models.

Additionally, we incorporate make- and segment-specific fixed effects to account for unobserved vehicle characteristics. We capture household-level unobserved heterogeneity through lognormal distributions on the coefficient of the income minus rental cost, negative lognormal distribution on the operating cost, and normal distribution on the taste of driving in the indirect utility. Whereas the first distributional assumption ensures a positive effect of income on VKT and the positive marginal utility of consuming all other goods, the second assumption imposes a negative effect of operating cost on the VKT.

We assume that all households have the same choice set (i.e., 234 models for China and 81 models for India). Since the data include new car buyers, the surveyed households encounter a decision of which car to buy conditional on having already decided to buy a new car. Due to this data limitation, we do not consider an outside good option.

---

[2] We use the market share of diesel and petrol in 2017 to compute a weighted average of respective fuel prices (Rampal, 2021). We use prices from the middle of the year in Delhi, the national capital (ZeeBiz, 2017).

## 5. Results and discussion

We first test the model's in-sample fit in terms of market share and annual VKT at the segment level. We plot the observed and predicted market shares and VKT for China in Figure A.1, and the same variables are plotted for India in Figure A.2 of the appendix. The results indicate that both variables are recovered well using the estimated model, with slight underestimation of VKT for SUV, MPV, and Luxury categories for India.

Parameter estimates (except alternative-specific constants) of the joint model for India and China are presented in Table 6. Most parameters are statistically significant at a 0.01 significance level, with a few exceptions, and the directions of effects are intuitive. Male with higher car ownership and larger families drive more in both countries. Car buyers from both countries with larger families prefer larger vehicles (indicated by a positive interaction effect of family size with engine displacement in India and volume in China). Coefficients of operating cost and income minus annual rent are used to derive short-run fuel price and income elasticity of annual VKT. Specifically, fuel price elasticity using equation 3 is $\alpha_i p_j$, which we compute at the mean of $\alpha_i$ and sales-weighted mean operating cost. Income elasticity is computed in the same way.

Table 6 shows that short-run fuel price elasticity of VKT, considering changes only in vehicle use and not vehicle choice in response to changing fuel price, is -0.18 for India and -0.28 for China. The earlier estimates by Chugh and Cropper (2017) are -0.68 and -0.93 for diesel and petrol car owners in India, but they use old data from 2010. As discussed earlier, fuel price regulations have experienced radical changes in India since then. We are not aware of short-run fuel price elasticity estimates of VKT for China, but Lin and Zeng (2013) find that short-run elasticity for fuel consumption is not statistically different from zero[3]. The short-run fuel price elasticities of VKT for the USA are also below -0.16 (in magnitude) (Goodwin et al., 2004).

We find that the short-run income elasticity of VKT is 0.14 for India and 0.12 for China. Our estimate for India is lower than the earlier-reported estimate of 0.28 by Chugh and Cropper (2017). An earlier study on the Chinese market by Lin and Zeng (2013) uses time-series data from 1998 to 2007 and find that the income elasticity of VKT is not statistically different from zero even in the intermediate run. Similar low income elasticity of VKT in short-run has also been obtained for the USA (see Goodwin et al., 2004 for a review).

---

[3] Earlier intermediate/long-run estimates of fuel price elasticity of VKT for China are [-0.88, -0.58] (Lin and Zeng, 2013) and -0.59 (Tan et al., 2019).

**Table 6:** Parameter estimates of the demand model and short-run elasticity of vehicle kilometres travelled.

|  | India | | China | |
|---|---|---|---|---|
|  | Estimates | Standard error | Estimates | Standard error |
| | Fixed parameters ($\gamma$) | | | |
| Age (year/100) | -0.33*** | 0.10 | | |
| Female? | -0.13*** | 0.040 | -0.063*** | 0.013 |
| Number of cars | 0.091*** | 0.022 | 0.035** | 0.016 |
| Family size | 0.035*** | 0.0070 | 0.14** | 0.080 |
| Length x width x height ($10^{-10}$ mm$^3$) | | | -0.0077* | 0.0057 |
| Family size x length x width x height ($10^{-10}$ mm$^3$) | | | 0.064*** | 0.016 |
| Family size x Engine displacement (litres) | 0.0028*** | 0.00038 | | |
| | Random parameters ($\alpha$ and $\beta$) | | | |
| *Mean* | | | | |
| Income - rent ($10^5$ CNY and $10^6$ INR) ($\beta$) | -1.68*** | 0.12 | -2.79*** | 0.10 |
| Operating cost per km (CNY or INR) ($\alpha$) | -3.83*** | 0.27 | -1.61*** | 0.13 |
| *Standard deviation* | | | | |
| Income - rent ( and $10^6$ INR) ($\beta$) | 0.45*** | 0.088 | 0.31*** | 0.088 |
| Operating cost per km (CNY/INR) ($\alpha$) | 1.29*** | 0.11 | 1.44*** | 0.063 |
| Taste of driving ($\sigma$) | 0.64*** | 0.011 | 0.40*** | 0.0071 |
| Scale factor ($\mu$) | 9.53*** | 0.90 | 1.59*** | 0.20 |
| | Short-run elasticity of vehicle kilometres travel (VKT) | | | |
| Income | 0.14*** | 0.0083 | 0.12*** | 0.0071 |
| Fuel price or operating cost | -0.18*** | 0.059 | -0.28*** | 0.049 |
| Number of observations | 7894 | | 8951 | |
| Number of alternative | 81 | | 234 | |
| Alternative-specific constants | 29 | | 89 | |
| Loglikelihood | -39242 | | -52806 | |

*p < 10%, **p < 5%, ***p < 1%

**Note:** This table presents full information maximum likelihood coefficient estimates, but alternative-specific constants (ASCs) are not shown. Mean and standard deviation of random parameters are presented for the underlying normal distributions. Family size is normalized by 10 in China data.

## 5.1 Long-run elasticity estimates through simulation

To estimate the long-run elasticity of fuel consumption relative to fuel price and income, we accounted for changes in both vehicle choice and use in response to the considered drivers. Specifically, we increase fuel price and income by 5% for all car buyers (ceteris paribus) and let consumers adjust their preference for car type and usage. The long-run fuel price elasticity of fuel consumption is -0.12 and -0.15 for India and China, respectively[4]. The ranges of the existing intermediate/long-run elasticities for India and China are [-0.39, -0.29] (Chugh and Cropper, 2017) and [-0.50, -0.20] (Lin and Zeng, 2013). Our estimates of the long-run income elasticity of fuel consumption for India and China are 0.15 and 0.13, respectively.

---

[4] Since we do not have an outside good option, long-run elasticity relative to variables that change for the entire vehicle fleet (i.e., fuel price and income) might be underestimated. Specifically, since the total number of cars remains the same in the system, the effects of change in fuel price/income on vehicle fleet composition are biased. Chugh and Cropper (2017) also acknowledge this limitation.

The existing income/expenditure elasticities for India and China are 0.35 (Chugh and Cropper, 2017) and [0.27, 1.24] (He et al., 2017), respectively.

We now present elasticity of fuel consumption relative to car-specific attributes. We consider two car attributes, car price and fuel economy and compute long-run elasticities at the segment level. To compute long-run own-price elasticity for each car segment, we increase purchase price of all models in that segment by 5% from the baseline and allow consumers to settle for their new preferred car and new VKT. We also compute the segment-level fuel consumption in case of no rebound (i.e., VKT) effect, i.e. assuming that average model-level VKT is not affected due to changes in purchase price. We follow the same procedure to find own-fuel-economy elasticity of fuel consumption (with and without rebound effect).

The ranges of the segment-level own-price elasticities of fuel consumption for India and China are [-2.07,-0.38], and [-3.47,-0.16], respectively (see Table 7 for details). The existing ranges of these estimates for India and China are [-1.86, -0.68] (Chugh and Cropper, 2017), and [−1.22, −0.46] (He et al., 2017), respectively. We observe that the effect of the car prices on fuel consumption is largely driven by the changes in the fleet composition as the segment-level market share elasticities are in sync with those of fuel consumption. In line with this observation, the purchase price has a low rebound effect in both markets – 0.96% for China and 1.40% for India. The sales-weighted mean of own-price elasticities of fuel consumption with and without rebound (VKT) effect are -0.651 and -0.642 for India, and -0.634 and -0.628 for China, respectively.

Table 8 shows the segment-level own-fuel-economy estimates of fuel consumption and VKT. Note that the increase in fuel economy brings three-dimensional effects on fuel consumption – change in unit operating cost, fleet composition, and VKT. If we consider all effects, a 1% increase in fuel economy would, on average, reduce fuel consumption by 0.571% in India and by 0.603% in China. However, if we ignore the third effect (i.e., VKT effect), these estimates are 0.689% for India and 0.743% for China. Thus, the rebound effect of fuel economy improvement on fuel consumption is around 17.1% for India and 18.8% for China. Such rebound effects cannot be captured in studies that do not model VKT with car preferences.

**Table 7:** Segment-level own-price elasticity

| Segments | Market share elasticity | Fuel consumption elasticity | Fuel consumption elasticity (no rebound) | Market share | Purchase Price ($10^5$ CNY and $10^6$ INR) |
|---|---|---|---|---|---|
| India | | | | | |
| Luxury | -1.98 | -2.07 | -2.00 | 0.4% | 4.89 |
| Upper Compact | -0.54 | -0.54 | -0.54 | 8.2% | 0.65 |
| MUV/ MPV | -0.82 | -0.84 | -0.83 | 8.6% | 1.09 |
| Entry Midsize | -0.65 | -0.66 | -0.65 | 12.9% | 0.76 |
| Entry Compact | -0.38 | -0.38 | -0.38 | 11.5% | 0.39 |
| Compact | -0.43 | -0.44 | -0.43 | 14.9% | 0.50 |
| Premium Compact | -0.68 | -0.69 | -0.68 | 16.7% | 0.74 |
| SUV | -0.84 | -0.87 | -0.85 | 15.8% | 1.19 |
| Premium SUV | -1.89 | -1.96 | -1.91 | 0.9% | 3.05 |
| Midsize | -0.82 | -0.84 | -0.82 | 4.8% | 1.09 |
| Premium Midsize | -0.88 | -0.91 | -0.89 | 0.3% | 1.92 |
| Van | -0.39 | -0.39 | -0.39 | 4.9% | 0.39 |
| **Sales-weighted average** | **-0.640** | **-0.651** | **-0.642** | | |
| China | | | | | |
| Compact | -0.23 | -0.24 | -0.23 | 1.6% | 0.53 |
| Compact Basic | -0.19 | -0.19 | -0.19 | 0.2% | 0.53 |
| Compact Luxury | -1.21 | -1.23 | -1.22 | 2.6% | 2.92 |
| Compact Luxury SUV | -1.23 | -1.24 | -1.23 | 1.4% | 3.13 |
| Compact Mini | -0.50 | -0.50 | -0.50 | 0.1% | 1.42 |
| Compact MPV | -0.30 | -0.30 | -0.30 | 5.2% | 0.63 |
| Compact SUV | -0.46 | -0.47 | -0.46 | 11.8% | 1.21 |
| Compact Upper | -0.37 | -0.38 | -0.37 | 2.9% | 0.85 |
| Large Luxury | -3.43 | -3.47 | -3.43 | 0.1% | 10.15 |
| Large Luxury SUV | -3.04 | -3.07 | -3.04 | 0.7% | 8.57 |
| Large MPV | -1.17 | -1.19 | -1.17 | 1.1% | 2.81 |
| Large SUV | -1.10 | -1.12 | -1.11 | 4.2% | 2.71 |
| Midsize | -0.46 | -0.46 | -0.46 | 20.5% | 1.21 |
| Midsize Basic | -0.34 | -0.35 | -0.34 | 10.6% | 0.85 |
| Midsize Luxury | -1.74 | -1.76 | -1.75 | 2.4% | 4.20 |
| Midsize Luxury SUV | -1.74 | -1.76 | -1.74 | 2.0% | 4.40 |
| Midsize MPV | -0.60 | -0.60 | -0.60 | 0.5% | 1.55 |
| Midsize SUV | -0.69 | -0.70 | -0.70 | 12.4% | 1.73 |
| Midsize Upper | -0.85 | -0.86 | -0.85 | 6.5% | 2.10 |
| Midsize Upper Economy | -0.65 | -0.65 | -0.65 | 5.0% | 1.54 |
| Mini Van | -0.16 | -0.16 | -0.16 | 0.8% | 0.37 |
| Small SUV | -0.34 | -0.34 | -0.34 | 7.6% | 0.81 |
| **Sales-weighted average** | **-0.626** | **-0.634** | **-0.628** | | |

**Note:** All elasticity estimates are statistically significant at a 0.01 significance level.

Table 8: Segment-level own-fuel economy elasticity

| Segments | Market share elasticity | Fuel consumption elasticity | Fuel consumption elasticity (no rebound) | VKT elasticity | Market share | Fuel economy (km/litre) |
|---|---|---|---|---|---|---|
| **India** | | | | | | |
| Luxury | 0.39 | -0.44 | -0.58 | 0.53 | 0.4% | 13.1 |
| Upper Compact | 0.30 | -0.55 | -0.67 | 0.42 | 8.2% | 16.2 |
| MUV/ MPV | 0.31 | -0.54 | -0.66 | 0.43 | 8.6% | 15.4 |
| Entry Midsize | 0.28 | -0.57 | -0.69 | 0.40 | 12.9% | 16.5 |
| Entry Compact | 0.26 | -0.59 | -0.70 | 0.38 | 11.5% | 17.6 |
| Compact | 0.26 | -0.59 | -0.70 | 0.38 | 14.9% | 17.0 |
| Premium Compact | 0.25 | -0.60 | -0.72 | 0.37 | 16.7% | 16.7 |
| SUV | 0.28 | -0.57 | -0.69 | 0.40 | 15.8% | 15.8 |
| Premium SUV | 0.38 | -0.45 | -0.58 | 0.53 | 0.9% | 13.1 |
| Midsize | 0.31 | -0.54 | -0.66 | 0.43 | 4.8% | 16.2 |
| Premium Midsize | 0.35 | -0.49 | -0.62 | 0.48 | 0.3% | 14.7 |
| Van | 0.32 | -0.52 | -0.64 | 0.45 | 4.9% | 15.2 |
| **Sales-weighted average** | **0.277** | **-0.571** | **-0.687** | **0.398** | | |
| **China** | | | | | | |
| Compact | 0.21 | -0.63 | -0.76 | 0.33 | 1.6% | 14.7 |
| Compact Basic | 0.21 | -0.63 | -0.75 | 0.34 | 0.2% | 14.6 |
| Compact Luxury | 0.25 | -0.56 | -0.71 | 0.41 | 2.6% | 11.3 |
| Compact Luxury SUV | 0.26 | -0.55 | -0.70 | 0.42 | 1.4% | 11.0 |
| Compact Mini | 0.21 | -0.62 | -0.75 | 0.34 | 0.1% | 14.4 |
| Compact MPV | 0.21 | -0.62 | -0.75 | 0.35 | 5.2% | 13.3 |
| Compact SUV | 0.22 | -0.60 | -0.74 | 0.37 | 11.8% | 12.0 |
| Compact Upper | 0.21 | -0.63 | -0.75 | 0.34 | 2.9% | 14.3 |
| Large Luxury | 0.29 | -0.52 | -0.68 | 0.45 | 0.1% | 10.1 |
| Large Luxury SUV | 0.31 | -0.49 | -0.66 | 0.49 | 0.7% | 9.1 |
| Large MPV | 0.30 | -0.50 | -0.67 | 0.47 | 1.1% | 9.5 |
| Large SUV | 0.28 | -0.53 | -0.69 | 0.44 | 4.2% | 10.0 |
| Midsize | 0.18 | -0.65 | -0.78 | 0.32 | 20.5% | 13.1 |
| Midsize Basic | 0.20 | -0.63 | -0.76 | 0.33 | 10.6% | 13.7 |
| Midsize Luxury | 0.27 | -0.54 | -0.69 | 0.43 | 2.4% | 10.5 |
| Midsize Luxury SUV | 0.28 | -0.52 | -0.68 | 0.45 | 2.0% | 10.0 |
| Midsize MPV | 0.26 | -0.55 | -0.70 | 0.42 | 0.5% | 11.3 |
| Midsize SUV | 0.23 | -0.58 | -0.73 | 0.39 | 12.4% | 11.1 |
| Midsize Upper | 0.25 | -0.57 | -0.72 | 0.40 | 6.5% | 11.3 |
| Midsize Upper Economy | 0.23 | -0.60 | -0.74 | 0.37 | 5.0% | 12.6 |
| Mini Van | 0.21 | -0.62 | -0.75 | 0.35 | 0.8% | 14.1 |
| Small SUV | 0.22 | -0.61 | -0.75 | 0.36 | 7.6% | 12.9 |
| **Sales-weighted average** | **0.220** | **-0.603** | **-0.743** | **0.366** | | |

**Note:** All elasticity estimates are statistically significant at a 0.01 significance level.

## 5.2 Assessment of a revenue neutral feebate policy

The estimated joint discrete-continuous model is used to predict the effect of a revenue-neutral feebate policy on the fleet fuel economy and fuel consumption in India and China. The specification and main results of the feebate policies are summarized in Table 9. We use sales-weighted average fleet fuel economy as the anchor point – 16.38 km/litre for India and 12.32 km/litre for China. To achieve revenue neutrality, we use a fee rate of $1846.2 per km/litre and a rebate of $1407.7 km/litre for India. The fee rate and rebate values are $1938.4 and $1627.7 km/litre for China. After converting fee rates into the proportion of retail price, we find that sales-weighted mean proportion of fees is much lower than that of rebates for China (7.41% vs. 14.56%) because car models with lower fuel economy belong to expensive luxury models.

The differences in total feebate and rebate are just $3000 for India and $10,000 for China, and thus feebate policies for both markets are approximately revenue-neutral. The revenue-neutral feebate policy leads to 0.812% and 0.795% improvement in sales-weighted fleet fuel economy for India and China, which translates into fuel savings of 0.703% and 0.688%, respectively. If the effect of change in car prices on VKT is ignored (i.e., rebound effect), the fuel savings are slightly higher – 0.758% for India and 0.699% for China. These results imply that the rebound effect associated with the feebate policy is 7.3% for India and 1.6% for China. Table 10 analyses the change in segment-level market shares due to the respective feebate policy. The results indicate that the feebate policy is working as expected, i.e. the market shares of segments with fuel economy above the anchor point increase due to the feebate policy. However, an increase in the actual market share of fuel-efficient vehicles due to the feebate policy is not substantial because their fuel economy is not too far from the anchor point. These results indicate that the structure of the fleet and low purchase price elasticity in both countries make the feebate less effective as a fuel conservation policy.

**Table 9:** The impact of a revenue neutral feebate policy

|  | India | China |
| --- | --- | --- |
| Anchor (km/litre) | 16.38 | 12.32 |
| Rebate (USD per km/litre) | 1407.7 | 1627.7 |
| Feebate (USD per km/litre) | 1846.2 | 1938.4 |
| Rebate (% retail price) | 15.37% | 14.56% |
| Feebate (% retail price) | 15.24% | 7.41% |
| Fuel economy increase (%) | 0.812% | 0.795% |
| Fuel savings (%) | 0.703% | 0.688% |
| Fuel savings (%, no rebound) | 0.758% | 0.699% |
| Original fuel consumption (litre) | 1.8980E+09 | 2.2819E+10 |
| Final fuel consumption (litre) | 1.8847E+09 | 2.2662E+10 |
| Final fuel consumption (no rebound, litre) | 1.8836E+09 | 2.2660E+10 |
| Total rebate (million USD) | 1.776E+03 | 2.0570E+04 |
| Total feebate (million USD) | 1.773E+03 | 2.0569E+04 |

Table 10: Segment-level effect of the feebate policy

| Segments | Market share change (%) | Fuel consumption change (%) | Fuel consumption change (%) (no rebound) | Fuel economy (km/litre) | Purchase Price ($10^5$ CNY and $10^6$ INR) | Market share |
|---|---|---|---|---|---|---|
| **India** | | | | | | |
| Luxury | -14.2 | -15.3 | -14.8 | 13.1 | 4.89 | 0.4% |
| Upper Compact | -2.2 | -2.4 | -2.4 | 16.2 | 0.65 | 8.2% |
| MUV/ MPV | -7.8 | -8.9 | -8.8 | 15.4 | 1.09 | 8.6% |
| Entry Midsize | **0.8** | 0.7 | 0.6 | **16.5** | 0.76 | 12.9% |
| Entry Compact | **11.6** | 11.6 | 11.3 | **17.6** | 0.39 | 11.5% |
| Compact | **5.8** | 5.7 | 5.6 | **17.0** | 0.50 | 14.9% |
| Premium Compact | **1.1** | 1.0 | 0.9 | **16.7** | 0.74 | 16.7% |
| SUV | -3.9 | -5.2 | -5.1 | 15.8 | 1.19 | 15.8% |
| Premium SUV | -21.4 | -23.1 | -22.6 | 13.1 | 3.05 | 0.9% |
| Midsize | -1.0 | -1.3 | -1.3 | 16.2 | 1.09 | 4.8% |
| Premium Midsize | -7.5 | -8.5 | -8.2 | 14.7 | 1.92 | 0.3% |
| Van | -14.1 | -14.1 | -14.1 | 15.2 | 0.39 | 4.9% |
| **China** | | | | | | |
| Compact | **13.2** | 13.1 | 13.0 | **14.7** | 0.53 | 1.6% |
| Compact Basic | **9.9** | 10.0 | 9.9 | **14.6** | 0.53 | 0.2% |
| Compact Luxury | -5.3 | -5.5 | -5.5 | 11.3 | 2.92 | 2.6% |
| Compact Luxury SUV | -6.4 | -6.5 | -6.5 | 11.0 | 3.13 | 1.4% |
| Compact Mini | **8.6** | 8.7 | 8.6 | **14.4** | 1.42 | 0.1% |
| Compact MPV | **5.7** | 5.8 | 5.7 | **13.3** | 0.63 | 5.2% |
| Compact SUV | -1.7 | -1.8 | -1.8 | 12.0 | 1.21 | 11.8% |
| Compact Upper | **10.5** | 10.5 | 10.4 | **14.3** | 0.85 | 2.9% |
| Large Luxury | -9.6 | -9.7 | -9.6 | 10.1 | 10.15 | 0.1% |
| Large Luxury SUV | -14.7 | -14.9 | -14.7 | 9.1 | 8.57 | 0.7% |
| Large MPV | -14.3 | -14.5 | -14.4 | 9.5 | 2.81 | 1.1% |
| Large SUV | -11.9 | -12.1 | -12.0 | 10.0 | 2.71 | 4.2% |
| Midsize | **4.5** | 4.5 | 4.4 | **13.1** | 1.21 | 20.5% |
| Midsize Basic | **7.4** | 7.4 | 7.3 | **13.7** | 0.85 | 10.6% |
| Midsize Luxury | -9.8 | -9.9 | -9.9 | 10.5 | 4.20 | 2.4% |
| Midsize Luxury SUV | -11.6 | -11.7 | -11.6 | 10.0 | 4.40 | 2.0% |
| Midsize MPV | -4.9 | -5.2 | -5.2 | 11.3 | 1.55 | 0.5% |
| Midsize SUV | -7.2 | -7.3 | -7.3 | 11.1 | 1.73 | 12.4% |
| Midsize Upper | -5.7 | -5.9 | -5.8 | 11.3 | 2.10 | 6.5% |
| Midsize Upper Economy | **1.3** | 1.2 | 1.2 | **12.6** | 1.54 | 5.0% |
| Mini Van | **9.1** | 9.2 | 9.1 | **14.1** | 0.37 | 0.8% |
| Small SUV | **3.4** | 3.2 | 3.2 | **12.9** | 0.81 | 7.6% |

**Note:** segments with fuel economy higher than pivot point (i.e., fleet average fuel economy) are in **bold**.

## 6. Conclusions

The effectiveness of a fuel conservation policy pivots around the sensitivity of consumers to purchase fuel-efficient cars and drive less. While several studies have considered both aspects in evaluating the fuel conservation policies in developed countries, limited evidence exists for developing countries like India and China. This study estimates a discrete-continuous structural econometric model to jointly analyse household-level data on vehicle preferences and vehicle kilometres travelled (VKT) of new car buyers from India and China who have purchased new cars in 2016-17. The model parameters are used to estimate various elasticities and rebound effects. A revenue-neutral feebate policy is also simulated to investigate its effectiveness in developing countries.

The short-run VKT elasticities relative to fuel price and income are -0.18 and 0.14 for India, and -0.28 and 0.12 for China, respectively. Similarly, long-run fuel price and income elasticities of fuel consumption are -0.12 and 0.15 for India, and -0.15 and 0.13 for China, respectively. Whereas the own-price elasticity of fuel consumption varies substantially across car segments, the own-fuel-economy elasticity is less heterogeneous. Sales-weighted direct (i.e., own) price and fuel economy elasticities of fuel consumption are -0.651 and -0.571 for India, and -0.634 and -0.603 for China, respectively. The purchase price has a low rebound effect in both markets, but the magnitude of fuel consumption elasticity relative to fuel economy reduces by 17.1% for India and 18.8% for China due to the rebound effect (i.e., induced travel caused by lower operating cost).

The revenue-neutral feebate policy could improve the sales-weighted mean fleet fuel economy of India and China by only 0.811% and 0.795%, with fuel savings of 0.701% and 0.688%, respectively. The negligible sensitivity of VKT relative to purchase price translates into a low rebound effect of the feebate policy, but it does not appear to be an effective fuel conservation policy in India and China.

Given the low responsiveness, none of the demand-side policies considered in this study appears effective in reducing fuel consumption, and thus carbon emissions, especially to the extent required to meet long-term temperature goals while maintaining new car sales. Low responsiveness combined with the need for deep decarbonization would result in either extremely high fuel and car taxes or the acceptance of lower economic output due to fewer new car sales. Given that fuel prices in both countries are already amongst the highest among top oil-consuming nations (Jacob, 2018), it is unclear whether there is much room for a further tax increase without risking public outcry, which could have serious political repercussions. Thus, policymakers in both countries have no choice but to implement supply-side policies, such as performance standards and mandates. Research suggests that such supply-side policies are less cost-effective than demand-side policies such as fuel taxes, which directly address the externality of reducing fuel consumption (Karplus et al., 2013). However, since the reasons for the increase in car prices as a result of such supply-side policies remain largely unknown to consumers, they are often considered more politically acceptable than government-enforced taxation.

We highlight that the results should be interpreted considering two caveats. First, we do not consider the outside good option because consumers in the sample encounter with a decision of "which car to buy?" instead of "whether to buy a car or not?" This restriction leads to the constant number of cars in the system, i.e. reduction or increment in car sales in elasticity calculations and feebate policy simulation could not be considered. That being said, our analysis is consistent with both developing countries' preference for policies that can reduce carbon emissions while sustaining new car sales, given their economic growth ambitions and the significant contribution of the new car market to each country's economic output. Second, the study does not consider the used car market due to the data limitations. Given that used car market is as big as the new car market in China (Yingying, 2021) and India (Saleem, 2019), future studies should find a way to account for interactions between used and new car markets.

# References

Bansal, P., Dua, R., Krueger, R., & Graham, D. J. (2021). Fuel economy valuation and preferences of Indian two-wheeler buyers. *Journal of Cleaner Production*, *294*, 126328.

Bansal, P., Keshavarzzadeh, V., Guevara, A., Daziano, R. A., & Li, S. (2021). Designed Quadrature to Approximate Integrals in Maximum Simulated Likelihood Estimation. *The Econometrics Journal*. DOI: https://doi.org/10.1093/ectj/utab023.

Bento, A. M., Goulder, L. H., Jacobsen, M. R., & Von Haefen, R. H. (2009). Distributional and efficiency impacts of increased US gasoline taxes. *American Economic Review*, *99*(3), 667-99.

Bhardwaj, D. (2020). Diesel Costs More Than Petrol, Should You Buy A Petrol or Diesel Car? URL: https://www.cars24.com/blog/diesel-costs-more-than-petrol-should-you-buy-a-petrol-or-diesel-car/

Chen, Y., Lawell, C. Y. C. L., & Wang, Y. (2020). The Chinese automobile industry and government policy. *Research in Transportation Economics*, *84*, 100849.

Chen, Y., & Lawell, C. Y. L. (2020). Fuel efficiency policies in the Chinese automobile market: Evidence from a random coefficients mixed oligopolistic differentiated products model. Working paper, Cornell University. URL: http://clinlawell.dyson.cornell.edu/China_auto_mkt_govt_policy_paper.pdf

Chugh, R., & Cropper, M. (2017). The welfare effects of fuel conservation policies in a dual-fuel car market: Evidence from India. *Journal of Environmental Economics and Management*, *86*, 244-261.

Chugh, R., Cropper, M., & Narain, U. (2011). The cost of fuel economy in the Indian passenger vehicle market. *Energy Policy*, *39*(11), 7174-7183.

Dahl, C. A. (2012). Measuring global gasoline and diesel price and income elasticities. *Energy Policy*, *41*, 2-13.

Dargay, J., Gately, D., & Sommer, M. (2007). Vehicle ownership and income growth, worldwide: 1960-2030. *The Energy Journal*, *28*(4).

Darido, G., Torres-Montoya, M., & Mehndiratta, S. (2014). Urban transport and $CO_2$ emissions: some evidence from Chinese cities. *Wiley Interdisciplinary Reviews: Energy and Environment*, *3*(2), 122-155.

Dua, R., Bhatt, Y., & Suneja D. (2021). What Policy Levers Could Address India's Automobile-Related Externalities? URL: https://www.kapsarc.org/research/publications/what-policy-levers-could-address-india%ca%bcs-automobile-related-externalities/

Dua, R., Hardman, S., Bhatt, Y., & Suneja D. (2021). Enablers and disablers to plug-in electric vehicle adoption in India: Insights from a survey of experts. *Energy Reports*, *7*, 3171-3188.

Dubin, J. A., & McFadden, D. L. (1984). An Econometric Analysis of Residential Electric Appliance Holdings and Consumption. *Econometrica*, *52*(2), 345-362.

EPI (2020). 2020 Environmental Performance Index Results. URL: https://epi.yale.edu/epi-results/2020/component/epi

Feng, Y., Fullerton, D., & Gan, L. (2013). Vehicle choices, miles driven, and pollution policies. *Journal of Regulatory Economics*, *44*(1), 4-29.

Frondel, M., Ritter, N., & Vance, C. (2012). Heterogeneity in the rebound effect: Further evidence for Germany. *Energy Economics*, *34*(2), 461-467.

Fullerton, D., Gan, L., & Hattori, M. (2015). A model to evaluate vehicle emission incentive policies in Japan. *Environmental Economics and Policy Studies*, *17*(1), 79-108.

Goldberg, P. K. (1998). The effects of the corporate average fuel efficiency standards in the US. *The Journal of Industrial Economics*, *46*(1), 1-33.

Gupta, S., Huddar, N., Iyer, B., & Möller, T. (2018). The future of mobility in India's passenger-vehicle market. URL: https://www.mckinsey.com/industries/automotive-and-assembly/our-insights/the-future-of-mobility-in-indias-passenger-vehicle-market

Goodwin, P., Dargay, J., & Hanly, M. (2004). Elasticities of road traffic and fuel consumption with respect to price and income: a review. *Transport Reviews*, *24*(3), 275-292.

Hao, H., Wang, H., & Yi, R. (2011a). Hybrid modeling of China's vehicle ownership and projection through 2050. *Energy*, *36*(2), 1351-1361.

Hao, H., Wang, H., Ouyang, M., & Cheng, F. (2011b). Vehicle survival patterns in China. *Science China Technological Sciences*, *54*(3), 625-629.

He, L. Y., Yang, S., & Chang, D. (2017). Oil price uncertainty, transport fuel demand and public health. *International Journal of Environmental Research and Public Health*, *14*(3), 245.

Jacob, S. (2018). Indian fuel prices fourth highest among top 10 crude oil consuming nations. URL: https://www.business-standard.com/article/economy-policy/indian-motor-fuel-prices-fourth-highest-among-top-10-crude-oil-nations-118052700624_1.html

Jacobsen, M. R. (2013). Evaluating US fuel economy standards in a model with producer and household heterogeneity. *American Economic Journal: Economic Policy*, *5*(2), 148-87.

Janssens-Maenhout, G., Crippa, M., Guizzardi, D., Muntean, M., Schaaf, E., Olivier, J. G., ... & Schure, K. M. (2017).*Fossil $CO_2$ & GHG Emissions of All World Countries* (Vol. 107877). Luxembourg: Publications Office of the European Union.

Karplus, V. J., Paltsev, S., Babiker, M., & Reilly, J. M. (2013). Should a vehicle fuel economy standard be combined with an economy-wide greenhouse gas emissions constraint? Implications for energy and climate policy in the United States. *Energy Economics*, *36*, 322-333.

Lin, C. Y. C., & Zeng, J. J. (2013). The elasticity of demand for gasoline in China. *Energy Policy*, *59*, 189-197.

Menon, B. G., & Mahanty, B. (2015). Assessing the effectiveness of alternative policies in conjunction with energy efficiency improvement policy in India. *Environmental Modeling & Assessment*, *20*(6), 609-624.

Niti Aayog and Rocky Mountain Institute (2017a). India leaps ahead: transformative mobility solutions for all. URL: https://rmi.org/insight/india-leaps-ahead-transformative-mobility-solutions-for-all/

Niti Aayog and Rocky Mountain Institute (2017b). Valuing society first: An assessment of the potential for a feebate policy in India. URL: https://rmi.org/insight/india-leaps-ahead-important-policy-lever-feebates/

Ou, S., Lin, Z., Xu, G., Hao, X., Li, H., Gao, Z., ... & Bouchard, J. (2020). The retailed gasoline price in China: Time-series analysis and future trend projection. *Energy*, *191*, 116544.

Rampal, N. (2021). From 58% to 17% market share: Why Indians don't like diesel cars anymore. URL: https://theprint.in/economy/from-58-to-17-market-share-why-indians-dont-like-diesel-cars-anymore/747710/

S&P Global (2019). An Overview of China's Auto Finance Market and Auto Loan Securitization.

Saleem, S.Z. (2019). Why the second-hand car market is seeing a boom. URL: https://www.livemint.com/auto-news/why-the-second-hand-car-market-is-seeing-a-boom-1557940318102.html

Sasi, A. (2019). Explained: The problem with diesel. URL: https://indianexpress.com/article/explained/diesel-cars-maruti-suzuki-bs-vi-standards-emission-5699556/

Sheldon, T. L., & Dua, R. (2020). Effectiveness of China's plug-in electric vehicle subsidy. *Energy Economics*, *88*, 104773.

Sheldon, T. L., & Dua, R. (2021). How responsive is Saudi new vehicle fleet fuel economy to fuel-and vehicle-price policy levers?. *Energy Economics*, *97*, 105026.

Small, K. A., & Van Dender, K. (2007). Fuel efficiency and motor vehicle travel: the declining rebound effect. *The Energy Journal*, *28*(1).

Sun, Q., Xu, L., & Yin, H. (2016). Energy pricing reform and energy efficiency in China: Evidence from the automobile market. *Resource and Energy Economics*, *44*, 39-51.

Tan, J., Xiao, J., & Zhou, X. (2019). Market equilibrium and welfare effects of a fuel tax in China: The impact of consumers' response through driving patterns. *Journal of Environmental Economics and Management*, *93*, 20-43.

Timperley, J. (2019). The Carbon Brief Profile: India. URL: https://www.carbonbrief.org/the-carbon-brief-profile-india

Wang, Y., Teter, J., & Sperling, D. (2011). China's soaring vehicle population: even greater than forecasted?. *Energy Policy*, *39*(6), 3296-3306.

Wang, J., Wu, Q., Liu, J., Yang, H., Yin, M., Chen, S., ... & Huang, Q. (2019). Vehicle emission and atmospheric pollution in China: problems, progress, and prospects. *PeerJ*, *7*, e6932.

West, S. E. (2004). Distributional effects of alternative vehicle pollution control policies. *Journal of Public Economics*, *88*(3-4), 735-757.

Xiao, J., & Ju, H. (2014). Market equilibrium and the environmental effects of tax adjustments in China's automobile industry. *Review of Economics and Statistics*, *96*(2), 306-317.

Yang, Z., & Tang, M. (2019). Welfare analysis of government subsidy programs for fuel-efficient vehicles and new energy vehicles in China. *Environmental and Resource Economics*, *74*(2), 911-937.

Yingying, C. (2021). Used car market sees steady sales growth.
URL: https://www.chinadailyhk.com/article/230273

ZeeBiz (2017). Petrol, diesel prices for July 7, 2017: We tell you city-wise rates. URL: https://www.zeebiz.com/india/news-petrol-diesel-prices-for-july-7-2017-we-tell-you-city-wise-rates-18296

# Appendix

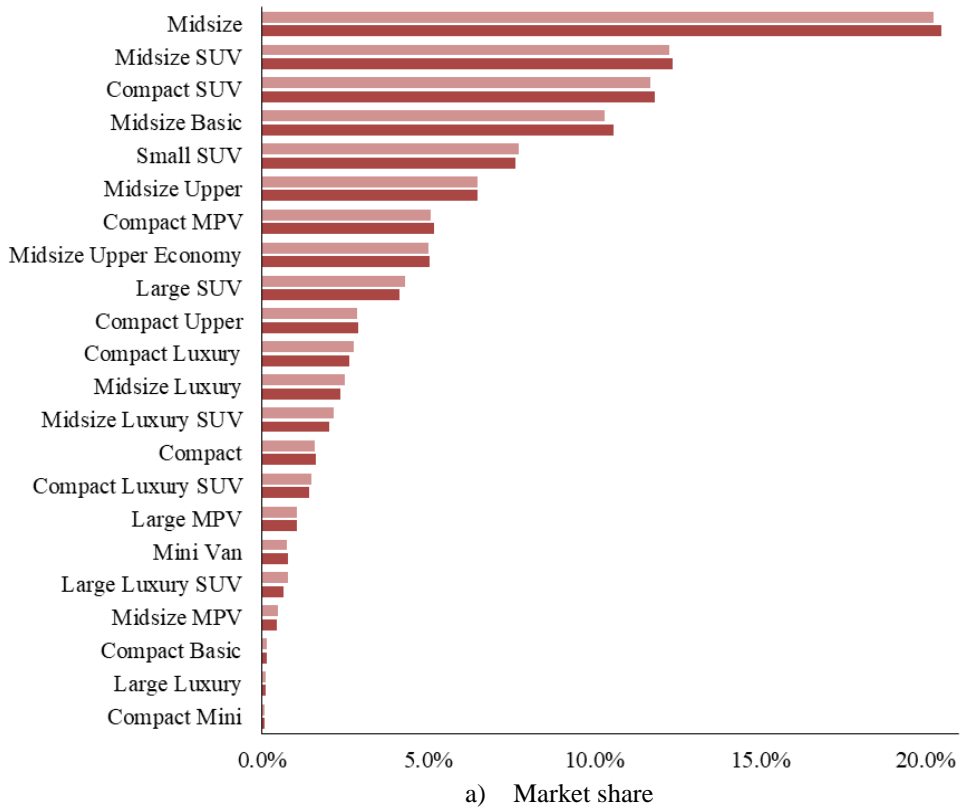

a) Market share

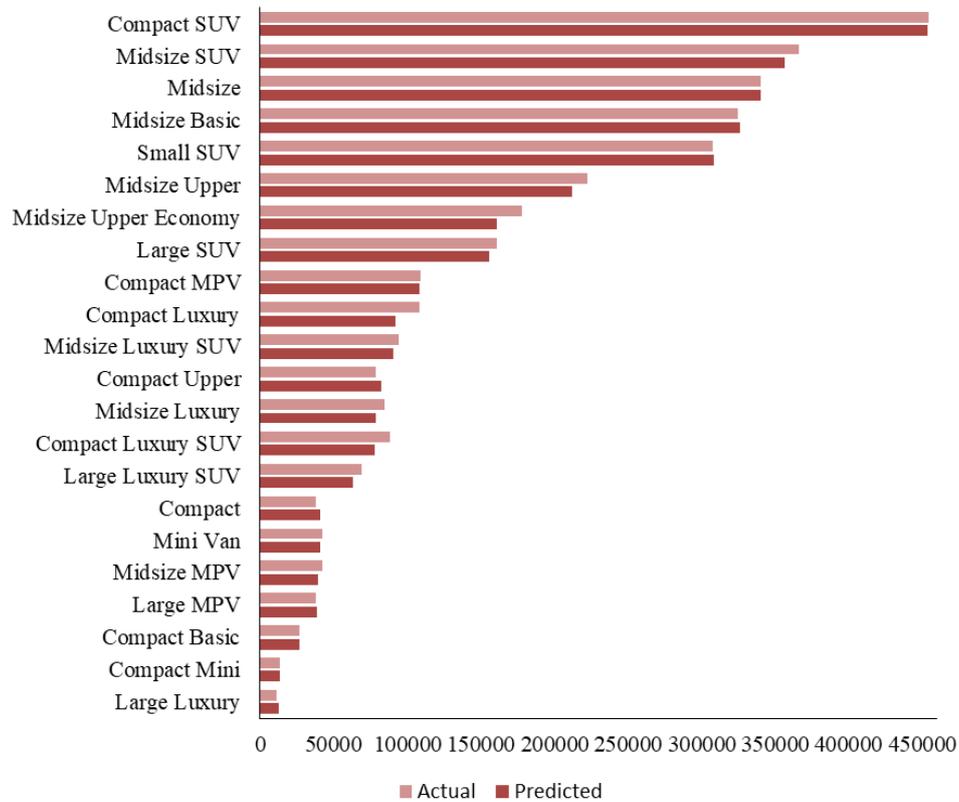

b) Vehicle kilometres travelled (sum of average of model-level VKT in a segment)

**Figure A.1:** Predicted market share and vehicle kilometres travelled (VKT) at the segment level in China.

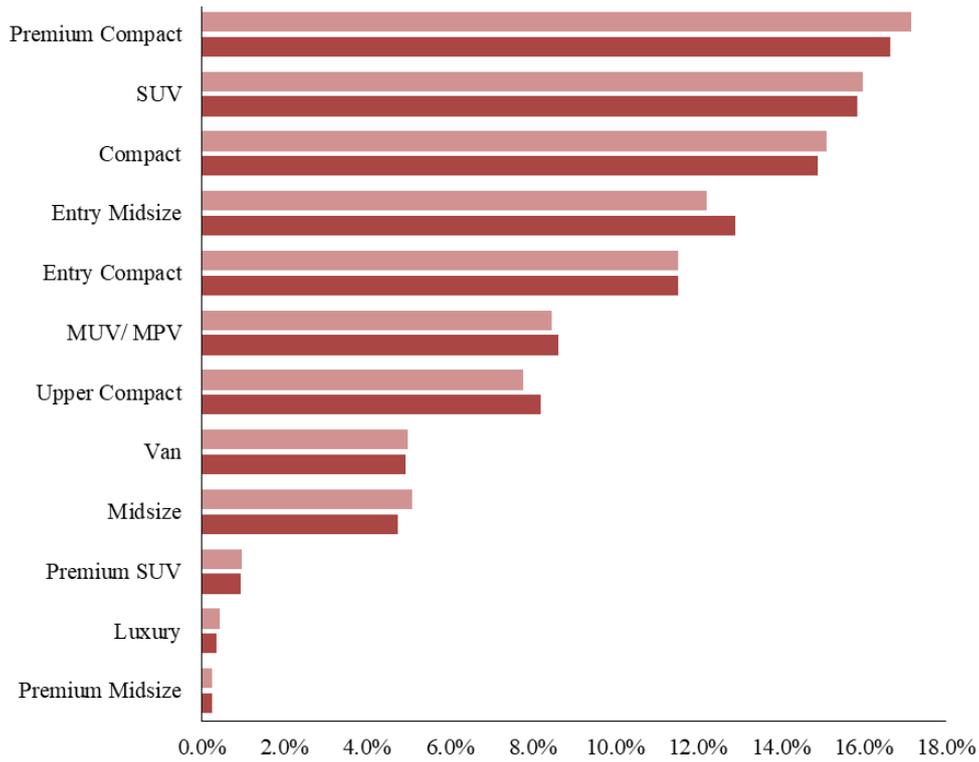

a) Market share

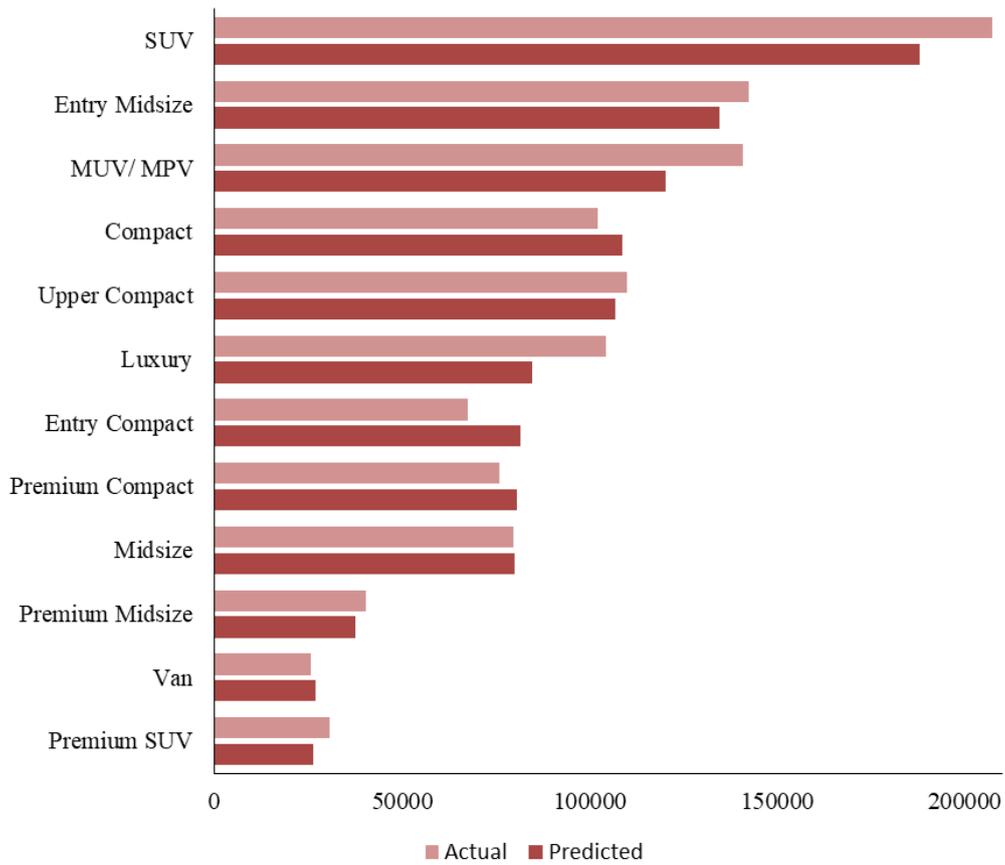

b) Vehicle kilometres travelled (sum of average of model-level VKT in a segment)

**Figure A.2:** Predicted market share and vehicle kilometres travelled (VKT) at the segment level in India.